\documentclass[sigconf]{acmart}

\AtBeginDocument{%
  \providecommand\BibTeX{{%
    \normalfont B\kern-0.5em{\scshape i\kern-0.25em b}\kern-0.8em\TeX}}}

\setcopyright{acmcopyright}
\copyrightyear{2020}
\acmYear{2020}
\acmDOI{10.1145/1122445.1122456}

\acmConference[UbiComp '20]{UbiComp '20}{September 12--16, 2020}{Cancun, Mexico}
\acmPrice{15.00}
\acmISBN{978-1-4503-XXXX-X/18/06}

\begin{document}

\title{TangToys: Smart Toys that can Communicate and Improve Children's Wellbeing}

\author{Kieran Woodward}
\email{kieran.woodward@ntu.ac.uk}
\affiliation{%
  \institution{Nottingham Trent University}
  \city{Nottingham}
  \country{UK}
}

\author{Eiman Kanjo}
\email{eiman.kanjo@ntu.ac.uk}
\affiliation{%
  \institution{Nottingham Trent University}
  \city{Nottingham}
  \country{UK}
}

\author{David J Brown}
\affiliation{%
  \institution{Nottingham Trent University}
  \city{Nottingham}
  \country{UK}
}

\author{Becky Inkster}
\affiliation{%
  \institution{University of Cambridge}
  \city{Cambridge}
  \country{UK}
}

\renewcommand{\shortauthors}{Woodward et al.}

\begin{abstract}
Children can find it challenging to communicate their emotions especially when experiencing mental health challenges. Technological solutions may help children communicate digitally and receive support from one another as advances in networking and sensors enable the real-time transmission of physical interactions. In this work, we pursue the design of multiple tangible user interfaces designed for children containing multiple sensors and feedback actuators. Bluetooth is used to provide communication between Tangible Toys (TangToys) enabling peer to peer support groups to be developed and allowing feedback to be issued whenever other children are nearby. TangToys can provide a non-intrusive means for children to communicate their wellbeing through play.
\end{abstract}



\maketitle
\section{Introduction}

The mental wellbeing of children is increasingly important as more young people than ever before are experiencing high levels of stress \cite{AmericanPsychologicalAssociation}. Tangible User Interfaces (TUIs) present new opportunities to digitise physical interfaces to help children communicate their wellbeing. Recent advances in microcontrollers and sensors enable small interfaces to be developed that can process and communicate sensor data in real-time. Children's toys represent an ideal embodiment for TUIs as they provide sufficient space for the electronics and encourage tactile interactions. Although a limited number of TUIs for mental wellbeing have previously been developed, many of these were not engaging for children and often contained physiological sensors which prevent physical interactions commonly used by children to interact with objects such as toys.

While there are many challenges in developing mental wellbeing interfaces, the decreasing cost and increasing capability of networking, sensors and microcontrollers is enabling new forms of interfaces to be developed \cite{Woodward2019}. An interface that could actively monitor and enable the communication of a user's physical interactions and mood would be beneficial for all. Through the use of Bluetooth Low Energy (BLE), TUIs can communicate with one another enabling real-time communication networks to be developed.

TUIs have previously been used to provide a method to communicate emotions and mental health states \cite{Woodward2018b}. Emoball \cite{Bravo2015} is a physical ball that enabled users to report their emotions by squeezing the device. Similarly, Subtle Stone \cite{Balaam2009} allowed users to express their emotions as a colour on the stone. Using colours to represent emotions enabled the private communication of emotions to only those who understood such colour representations. Mood TUI \cite{Sarzotti2018} also enabled users to self report their emotions but additionally collected data from the users’ smartphones such as location and heart rate. Overall, participants in this study found TUIs exciting to use, and that a small sized device was key for sustained interactions.

The vast majority of previously developed wellbeing interfaces have utilised self-reporting, which children in particular may find challenging. Recent developments in non-invasive sensors introduce the possibility to objectively and intuitively measure physical interactions and physiological changes in real-time. Motion data collected through accelerometers, gyroscopes and magnetometers could be used to measure physical interactions in addition to physiological sensors to monitor indicators of wellbeing. Previously, motion data has been used to infer emotions with 81.2\% accuracy across 3 classes \cite{Zhang2016}. However, similar studies reported lower levels of accuracy between 50\% - 72\% \cite{Quiroz2017} \cite{Olsen2017} \cite{Hossain2014} when inferring emotions from motion data.
 
The ability to measure and communicate wellbeing through non-invasive sensors presents many opportunities. This research introduces Tangible Toys (TangToys) with the aim of communicating mental wellbeing inferred from the embedded sensors. Initial prototypes embed sensors used to measure interactions and well-being and Bluetooth to enable real-time communication in-situ settings. The ability for devices to communicate with one another enables friends to communicate when socially distant and the ability to discover other nearby users. Furthermore, this work highlights key directions for the continued refinement of TangToys.

\section{Tangible Toys (TangToys)}
Few sensor based interfaces have been designed for children even though children traditionally find it challenging to communicate their mental wellbeing \cite{England2016}. We introduce the concept of TangToys as children's toys that embed electronics to measure tangible interactions. The interfaces can vary in shape, size and material ranging from soft balls and teddies designed for younger children, to 3D printed fidgeting cubes designed for older children. As children physically interact with TangToys in the same way as traditional toys all of the interfaces are suitable for children and encourage engagement by resembling similar toys.
\subsection{Communication Framework}
Embedding sensors within toys that can communicate with one another through Bluetooth offers many new opportunities for real-time interactions. BLE 4.2 has a range of around 50m allowing TangToys to communicate with one another in locations such as playgrounds. In the following sections we present two opportunities for real-time digital social interaction between TangToys.
\begin{table}[!h]
  \caption{Initial TangToys prototypes.}
  \label{tab:interfaces}
  \begin{tabular}{|p{1cm}|p{2.5cm}|p{4cm}|}
    \toprule
   Device&Image&Description\\
    \midrule
    Ball &\vspace{0cm} \includegraphics[width=20mm]{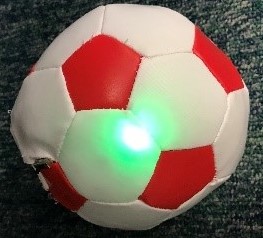} & A soft ball embedding 9-DOF IMU to measure motion, capacitive sensors to measure touch and Multi-coloured LEDs to perform visual feedback. \\
    \hline
    Cube &\vspace{0cm} \includegraphics[width=20mm]{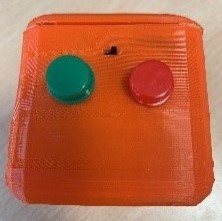} & A 3D printed cube embedding 9-DOF IMU, capacitive touch, HR, EDA sensors and haptic feedback\\
    \hline
    Teddy &\vspace{0cm} \includegraphics[width=20mm]{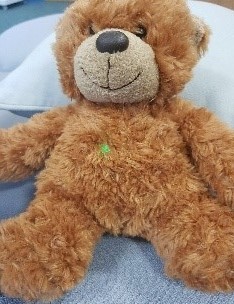} & A soft teddy embedding 9-DOF IMU, capacitive touch sensor and visual feedback\\
    \hline
    Torus  &\vspace{0cm} \includegraphics[width=20mm]{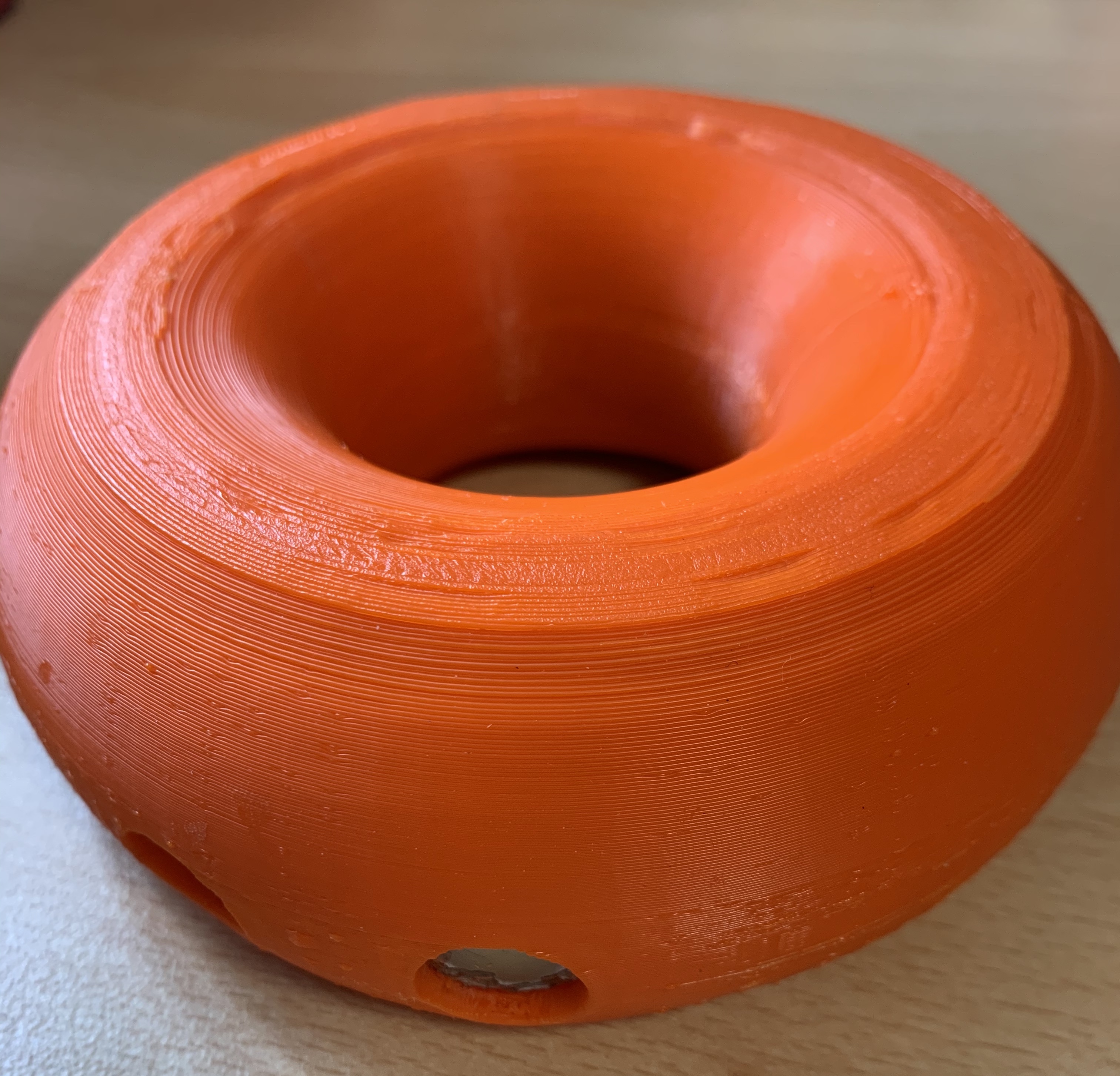} & A 3D printed tours embedding HR, EDA, 9-DOF IMU, capacitive touch sensor and haptic feedback\\
        \hline
    Teddy  &\vspace{0cm} \includegraphics[width=20mm]{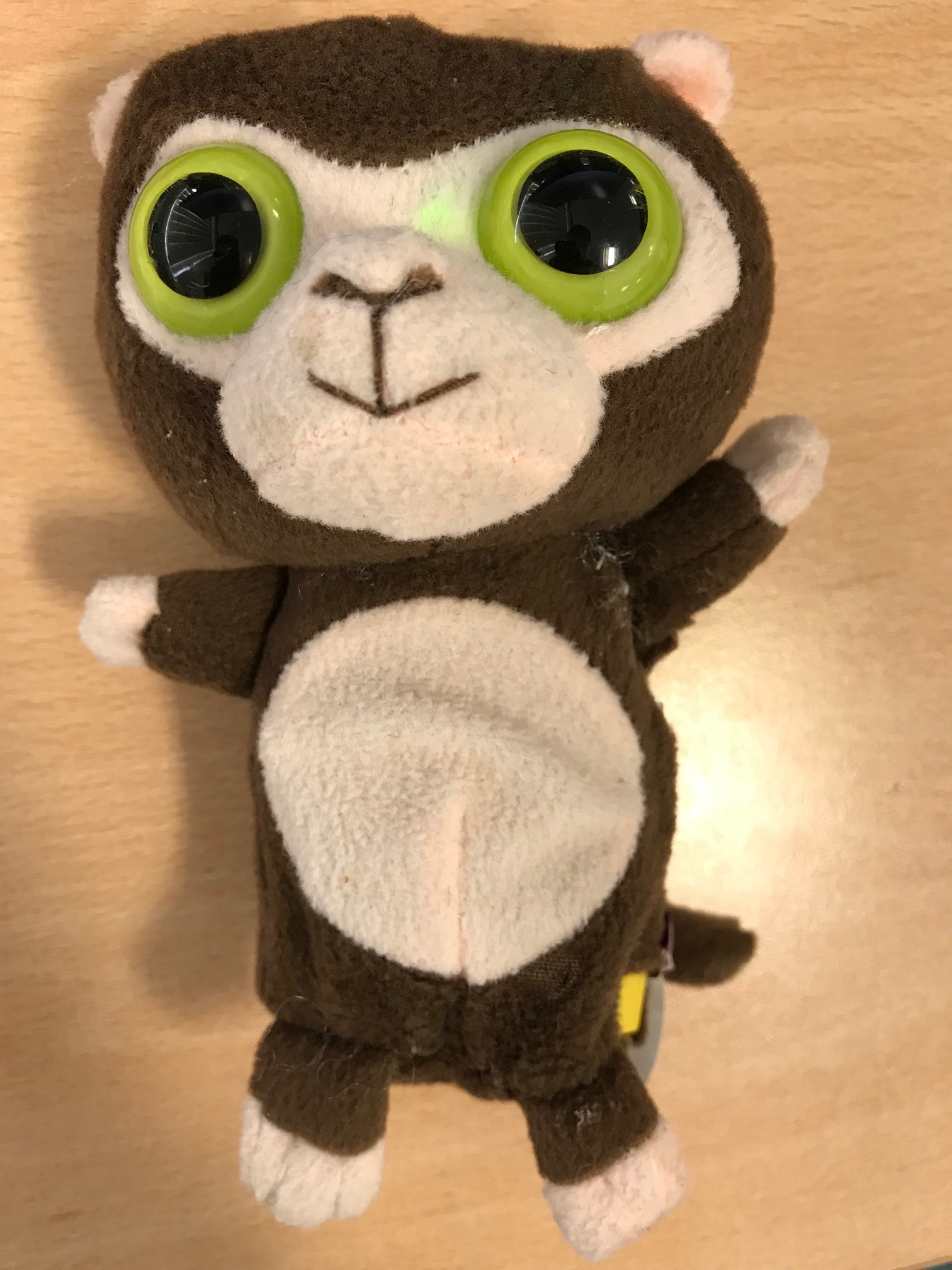} & A soft teddy embedding 9-DOF IMU, capacitive touch sensor, haptic and visual feedback\\
    
  \bottomrule
\end{tabular}
\end{table}

Table \ref{tab:interfaces} shows the five TangToys developed during a co-design and co-creation workshop including 2 soft teddies, a soft ball and a 3D printed cube and torus. Each TangToy includes a microcontroller and micro SD card to record all interactions along with bluetooth 4.2 for communication. A range of sensors can be used to monitor children's interactions with the toys including capacitive sensors to measure touch and 9-Degree Of Freedom Inertial Measurement Unit (9-DOF IMU) to measure motion. Physiological sensors can also be embedded within the toys such as Heart Rate (HR) sensors as they directly correlated with the sympathetic nervous system helping to monitor mental wellbeing \cite{Sharma2012} \cite{Alajmi2013}. All of the TangToys measure motion and touch interactions while only the 3D printed interfaces designed for older children include HR sensors to measure physiological changes.

In addition to the sensors, TangToys can provide real-time interventional feedback \cite{Woodward2018a}. Haptic feedback to provide the sensation of touch has been included within some of the developed prototypes. Haptic feedback provides a physical sense resembling touch which can improve mental wellbeing \cite{Klamet2016} \cite{Corbett2016}. Additionally, visual feedback in the form of multi-coloured LEDs has been included within the soft ball and teddy prototypes. These forms of feedback can function as real-time interventions if poor mental wellbeing can be detected, helping to alert the user.

TangToys have been presented in focus groups to teachers of young students with mild to moderate learning disabilities to provide feedback on the design and functionality of the interfaces \cite{Woodward2019a}. Teachers considered the methods used to interact with TangToys suitable for children and believed the way in which children interact with the toys will indicate their wellbeing. Additionally, teachers liked the design of the toys as they appear similar to other toys helping to reduce stigma. Overall, the teachers reported the design, sensors and communication capabilities were all suitable for children and believed would promote the communication of wellbeing between friends.

\subsubsection{Peer to Peer Communication (P2P)}
\begin{figure}[h]
\centering
\includegraphics[width=7cm]{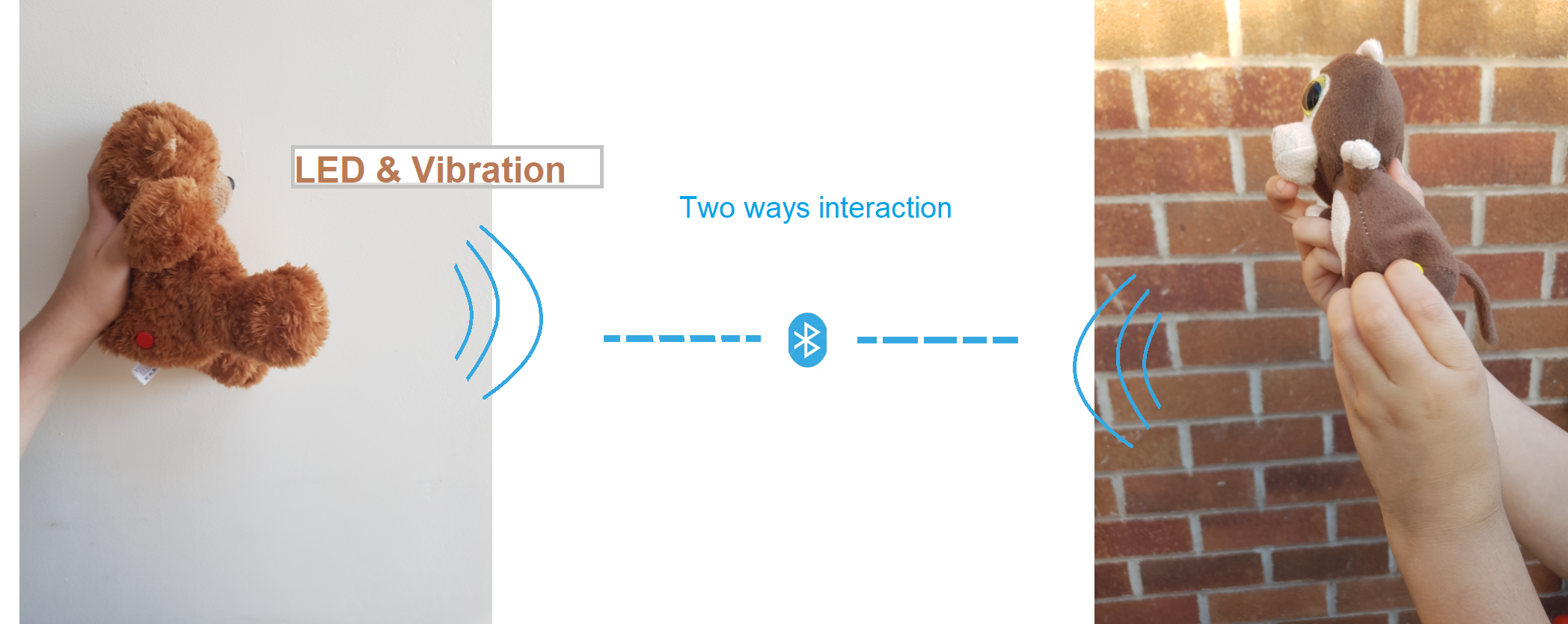}
\caption{Two children playing using TangToys.}
\label{Fig:Tnagtoyschildren}
\end{figure}

By utilising P2P communication it is possible for two connected devices to communicate with one another. This method of communication helps friends who may be nearby but socially distanced to provide physical communication that is not possible with other devices. When one user touches or moves their toy, the paired interface can react through the embedded visual or haptic feedback, providing a sense of physical interaction. The visual feedback and haptic patterns can differ dependent on the way in which the partner device is interacted with. For example, if a child is aggressively shaking their TangToy or touching it harshly this can result in prolonged sharp haptic feedback patterns being played on the paired device and red visual feedback. This enables friends to physically communicate how they are feeling and provide comfort to one another (see figure \ref{Fig:Tnagtoyschildren}).

The range of feedback that can be offered allows for emotions to be wirelessly communicated with haptic feedback providing a sense of presence as it simulates touch. Therefore, capacitive sensor data measuring touch can be actuated on the partner device using haptic feedback to simulate physical communication. Furthermore, haptic feedback can be used to comfort as previous work has shows the potential of haptic feedback to improve wellbeing \cite{Kelling2016}, \cite{Azevedo2017}.

\subsubsection{Wireless Scanning}
Each TangToy can also use its Bluetooth capabilities to broadcast its presence to other TangToys. When a TangToy detects another device nearby this can initiate feedback being issued to alert the child of other nearby children. This allows a child to find other children who may require support when not near their friends to facilitate peer to peer communication. These children can then interact with the devices to form a support group to communicate their wellbeing to each other. The feedback actuated when detecting other devices can be impacted by the number of nearby interfaces. For example, if a single child is detected nearby more subtle haptic feedback can be issued compared with more pronounced feedback when multiple children are nearby. Similar, the colour displayed on the TangToy can change dependent on the number of users located nearby to alert the user visually. Using this method of interaction would not enable the same capabilities as the P2P communication, but would enable each device to interact automatically with other nearby devices, and afford a sense of 'togetherness'.

\section{Conclusion and Future Work}
We have presented TangToys, a new concept to combine tangible user interfaces with traditional toys. Various sensors can be embedded within TangToys to communicate physical interactions such as movement and touch in real-time with other TangToys. Using a peer to peer communication system enables friends to communicate their wellbeing through device interactions that can then be actuated using haptic and visual feedback on a friend's device. Alternatively, TangToys can simultaneously broadcast and scan for nearby devices allowing for TangToys to discover other TangToys and create local support networks.

In the future, TangToys should be trialled with children, potentially in schools where they will be able to communicate with each other through touch and haptic feedback. The impact of the communication networks can be measured along with the duration in which children use the interfaces. Parental monitoring could also be included through the use of a mobile app enabling parents to view previous interactions with the toys.
\section{Acknowledgement}
{\footnotesize This project has been funded by the Nurture Network (eNurture). eNurture is funded by UK Research and Innovation (UKRI) and their support is gratefully acknowledged (Grant reference: ES/S004467/1). Any views expressed here are those of the project investigators and do not necessarily represent the views of eNurture or UKRI.}

\bibliographystyle{ACM-Reference-Format}
\bibliography{sample-base}

\end{document}